\documentclass[12pt]{iopart}
\usepackage{graphicx}
\usepackage{cite}
\usepackage{amsfonts}
\usepackage{wrapfig}
\usepackage{srcltx}
\usepackage{color}
\def\tr{\mbox{tr}}
\def\id{{\mathbf 1}}
\def\sub#1{{\mbox{\tiny #1}}}
\def\headline#1{{\noindent\textbf{#1}: \\}}

\begin{document}

\title[Entanglement versus mutual information in quantum spin chains]
      {Entanglement versus mutual information in quantum spin chains}
\author{Jaegon Um$^1$, Hyunggyu Park$^1$, and Haye Hinrichsen$^2$}
\address{$^1$ School of Physics, Korea Institute for Advanced Study, Seoul 130-722, Korea}
\address{$^2$ Universit\"at W\"urzburg, Fakult\"at f\"ur Physik und Astronomie, \\ \ \ \ 97074 W\"urzburg, Germany}

%\vspace{-6mm}
\ead{slung@kias.re.kr}
\vspace{-2mm}
\ead{hgpark@kias.re.kr}
\vspace{-2mm}
\ead{hinrichsen@physik.uni-wuerzburg.de}

\begin{abstract}
The quantum entanglement $E$ of a bipartite quantum Ising chain is compared with the mutual information $I$ between the two parts after a local measurement of the classical spin configuration. As the model is conformally invariant, the entanglement measured in its ground state at the critical point is known to obey a certain scaling form. Surprisingly, the mutual information of classical spin configurations is found to obey the same scaling form, although with a different prefactor. Moreover, we find that mutual information and the entanglement obey the inequality $I\leq E$ in the ground state as well as in a dynamically evolving situation. This inequality holds for general bipartite systems in a pure state and can be proven using similar techniques as for Holevo's bound.      
\\
\\
\noindent{\bf Keywords: } 
Spin chains, ladders and planes (Theory), Entanglement in extended quantum systems (Theory)
\end{abstract}
%
%\pacs{03.67.-a, 03.67.Mn, 05.30.Rt}
%03.67.-a : Quantum information
%03.67.Mn : Entanglement measures, witnesses, and other characterizations
%05.30.Rt : Quantum phase transitions
\maketitle
%\parskip 2mm 

%==========================================================================
\section{Introduction}
%==========================================================================

\begin{figure}[t]
\centering\includegraphics[width=100mm]{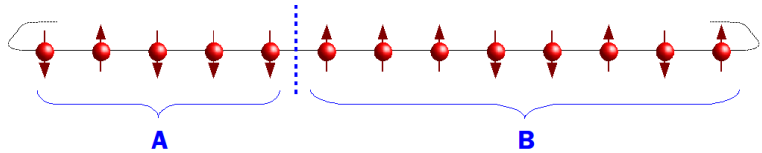}
\caption{Quantum spin chain with periodic boundary conditions. A measurement of spin orientations yields a classical spin configuration, as indicated by the arrows. Dividing the system into two fictitious parts $A$ and $B$, the respective spin configurations will be correlated, expressed in terms of the mutual information $I_\sub{A:B}$. \label{spinchain}}
\end{figure}

Recently the study of quantum spin chains from the perspective of quantum information theory attracted considerable attention. This applies in particular to entanglement studies of quantum spin chains in their ground state $\rho=\vert\psi_0\rangle\langle\psi_0\vert$. In these studies a quantum chain is fictitiously divided into two parts $A$ and $B$ (see Fig.~\ref{spinchain}). The quantum entanglement between the two parts is then given by 
\begin{equation}
E_\sub{A:B} \;=\; S[\rho_{A}] \;=\; S[\rho_{B}]\,, 
\label{eq1}
\end{equation}
where $S[\rho_{A,B}]=-\tr \left[ \rho_{A,B}\ln\rho_{A,B} \right]$ denotes the von-Neumann entropy of the reduced density matrices $\rho_{A}=\tr_{B}[\rho]$ and $\rho_{B}=\tr_{A}[\rho]$. The entanglement $E_\sub{A:B}$ is particularly interesting to study in the context of critical quantum chains with an underlying conformal symmetry~\cite{CI}, which are characterized by long-range correlations in the ground state. Using methods of conformal field theory it was shown in \cite{Vidal,Holzhey,Calabrese1, Calabrese2, Calabrese3} that the entanglement in such systems obeys the scaling form
\begin{equation}
\label{entanglement}
E_\sub{A:B} = a + \frac{c}{3} \ln \left( L\,f\Bigl(\frac{l}{L}\Bigr) \right)\,,
\end{equation}
where $L$ is the total length of the chain, $l$ is the length of section $A$, and $f$ is a scaling function. In this expression the constant $a$ is non-universal, meaning that it depends on the specific microscopic realization of the respective model, while the constant $c$ turns out to be universal. Remarkably, $c$ is equal to the so-called central charge of the underlying conformal field theory which labels the universality class. For example, the Ising universality class is characterized by the central charge $c=1/2$. 

%==================================
%==================================

To determine the entanglement experimentally one has to perform a variety of repeated measurements acting simultaneously on all spins in one of the sectors. Such as task is usually difficult to perform. In fact, it would be much simpler to measure individual spin orientations locally and to study the resulting classical spin configurations. Such a measurement is expected to destroy the existing entanglement and to convert it to some extent into classical correlations between the two parts of the system. Such classical correlations are usually quantified by the mutual information
\begin{equation}
\label{mutual}
I_\sub{A:B} \;=\; H_{A} + H_{B} - H_{AB}\,,
\end{equation}
where the terms on the right hand side denote the Shannon entropy of the classical spin configurations after the measurement in the sections $A,B$ and in the whole chain.

In this paper we address the question to what extent the quantum entanglement and the classical mutual information are related to each other. The main results are:
\begin{itemize}
\item In the quantum Ising model the mutual information obeys the same scaling law~(\ref{entanglement}) as the entanglement at the critical point, although with a different prefactor.
\item The mutual information obeys the inequality $I_\sub{A:B} \leq E_\sub{A:B}$ for general bipartite systems in a pure state.
\end{itemize}
The paper is organized as follows. In the following Section we first discuss 
the example of the quantum Ising chain, determining the entanglement and the 
mutual information in finite-size systems by different methods. In Sect. 3 we 
give a general proof of the inequality $I_\sub{A:B}\leq E_\sub{A:B}$. 
Finally,  in Sect. 4 we present a summary and discuss about the relation between our 
inequality evoked by local measurements and the monotonicity of the quantum 
relative entropy. 

%==========================================================================
\section{Entanglement and mutual information in the quantum Ising chain}
%==========================================================================

%==================================
\begin{figure}[t]
\centering\includegraphics[width=100mm]{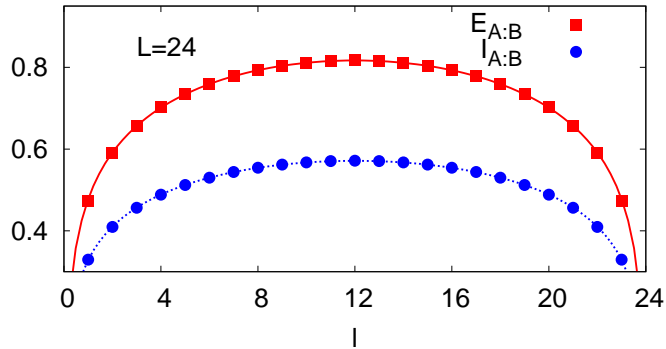}
\caption{Entanglement $E_\sub{A:B}(L,l)$ (squares) and mutual information $I_\sub{A:B}(L,l)$ (circles) in the ground state of the quantum Ising model with $L=24$ sites and periodic boundary condition at the critical point $g=1$. The solid and dashed line show the scaling functions according to the scaling forms (\ref{EScaling}) and (\ref{IScaling}), respectively. \label{scalingfunction}}
\end{figure}
%==================================

The quantum Ising chain with the length $L$ is defined by the Hamiltonian
\begin{equation}
H \;=\; -\sum_{i=1}^L \sigma_i^z \sigma_{i+1}^z  \,-\, g\sum_{i=1}^L \sigma^x_i\,,
\label{Ham:Ising}
\end{equation}
where $\sigma^{x,y,z}$ are Pauli matrices, $g$ is the strength of the 
transverse field. We use periodic boundary conditions by setting $L+1\equiv1$. The 
ground state of this model exhibits an order-disorder phase transition at the 
critical point $g_c=1$: At $g=0$ all spins are aligned in one direction of the 
$z$ axis, and as $g$ is increased the magnetization is weakened and vanishes 
at $g=1$, which means that the system goes to the paramagnetic phase in the 
$z$ direction at $g > 1$.

\headline{Scaling behavior $E_\sub{A:B}$ and $I_\sub{A:B}$ in the ground state}
%----------------------------------------------
%
To obtain the mutual information of classical spin configurations in the ground 
state, we have to compute the probability for each classical configuration after a measurement of 
local spin orientations. If we choose the $z$ axis for the spin orientations,
measuring $\sigma^z_1, \sigma^z_2,\ldots, \sigma^z_L$, these probabilities can be
computed from the coefficients of the ground state wave function 
$\vert \psi_{0} \rangle$ represented in the product basis of the eigenvectors 
of $\sigma_{i}^{z}$. For example, if we consider an Ising chain with two spins
in the ground state
\begin{equation}
\vert \psi_{0} \rangle = a_{1} \vert \uparrow \uparrow \rangle + 
a_{2} \vert \uparrow \downarrow \rangle + 
a_{3} \vert \downarrow \uparrow \rangle + 
a_{4} \vert \downarrow \downarrow \rangle ,
\end{equation}  
where $ \vert \uparrow \downarrow \rangle 
\equiv \vert \uparrow \rangle_1 \otimes \vert \downarrow \rangle_2$ is the 
product basis of the eigenvectors
$\vert \uparrow \rangle_i$ and $\vert \downarrow \rangle_i$ of 
$\sigma^{z}_{i}$, the probability $P(\{\sigma^{z}\})$ 
to obtain the classical configuration $\{\sigma^{z}\}$ 
is given by $P(\uparrow \uparrow) = a^2_{1}\,, P(\uparrow \downarrow )=a_{2}^2\,,  
P(\downarrow \uparrow )=a_{3}^2$, and $P(\downarrow \downarrow )=a_{4}^2$. 
To determine the ground state of the Hamiltonian~(\ref{Ham:Ising}) at finite $L$, 
we use the Lanczos method~\cite{lanczos}.  

Each classical configuration obtained from the measurement 
is now divided into two segments, $\{\sigma^{z} \} =
\{ \sigma^{z}_{A}, \sigma^{z}_{B} \}$.   
Determining the Shannon entropy of the probability distribution
in the segments and in the whole chain, 
\begin{eqnarray}
H_{A} &=& -\sum_{\{\sigma^{z}_{A}\}} P(\{\sigma^{z}_{A}\} )\ln P(\{\sigma^{z}_{A}\})\,,  \\
H_{B} &=& -\sum_{\{\sigma^{z}_{B}\}} P(\{\sigma^{z}_{B}\} )\ln P(\{\sigma^{z}_{B}\})\,,   \\
H_{AB} &=& -\sum_{\{\sigma^{z}_{A}, \sigma^{z}_{B}\}} P(\{\sigma^{z}_{A}, \sigma^{z}_{B}\} )\ln P(\{\sigma^{z}_{A}, \sigma^{z}_{B}\})\,, 
\end{eqnarray}
where $P(\{\sigma^{z}_{A} \}) =\sum_{\{ \sigma_{B}^{z} \}}
P( \{\sigma^{z}_{A}, \sigma^{z}_{B} \} )$ and
 $P(\{\sigma^{z}_{B} \}) =\sum_{\{\sigma_{A}^{z} \}}
P( \{\sigma^{z}_{A}, \sigma^{z}_{B} \} )$,
one can obtain the classical mutual information $I_\sub{A:B}(L,l)=
H_{A}+H_{B}-H_{AB}$ for a given length~$l$ of section $A$. 

On the other hand, to compute the entanglement $E_\sub{A:B}(L,l)$ 
between sections $A$~and~$B$, we perform the partial trace on the density matrix 
$\rho= \vert \psi_{0} \rangle \langle \psi_{0} \vert$ obtained from the
Lanczos method, leading to the reduced 
density matrices $\rho_{A}$ and $\rho_{B}$. Since the entanglement of a 
pure state is given by the von-Neumann entropy of the reduced density matrices 
(see Eq.~(\ref{eq1})), we can determine the entanglement $E_\sub{A:B}(L,l)$
by numerically diagonalizing  $\rho_{A}$ or $\rho_{B}$.  

Using this method, we calculate $E_\sub{A:B}(L,l)$ and $I_\sub{A:B}(L,l)$ 
for various values of $g$, up to $L=24$. As expected, at the critical point 
$g=1$ the entanglement $E_\sub{A:B}(L,l)$ is 
found to obey the scaling form~(\ref{entanglement}) as shown in 
Fig.~\ref{scalingfunction} with $f(\xi)=\frac{1}{\pi}\sin(\pi\xi)$. 
Surprisingly, the mutual information $I_\sub{A:B}(L,l)$ also obeys the same 
type of scaling form. To illustrate this finding, we plot two scaling forms
\begin{eqnarray}
\label{EScaling}
E_\sub{A:B}(L,l)        &=& a+ \frac16 \ln\Bigl(\frac{L}{\pi}\sin\frac{l\pi}{L}\Bigr) ,\\[2mm]
\label{IScaling}
I_\sub{A:B}(L,l)  &=& a' + \frac{c'}6 \ln\Bigl(\frac{L}{\pi}\sin\frac{l\pi}{L}\Bigr)  ,
\end{eqnarray}
as a solid and a dashed line in Fig.~\ref{scalingfunction}, together with the numerical data obtained by the Lanczos method. Here, we have used $a \approx 0.478$, $a' \approx 0.329$, and $c' \approx 0.715$. This means that the initial entanglement and the mutual information after the measurement differ only by a factor at the critical point. 

Note  that the classical mutual information after the measurement is always smaller than or equal to the initial entanglement in the ground state. The same observation holds in the off-critical case $g \neq 1$ (not shown here). \\

%==================================
\begin{figure}[t]
\centering\includegraphics[width=140mm]{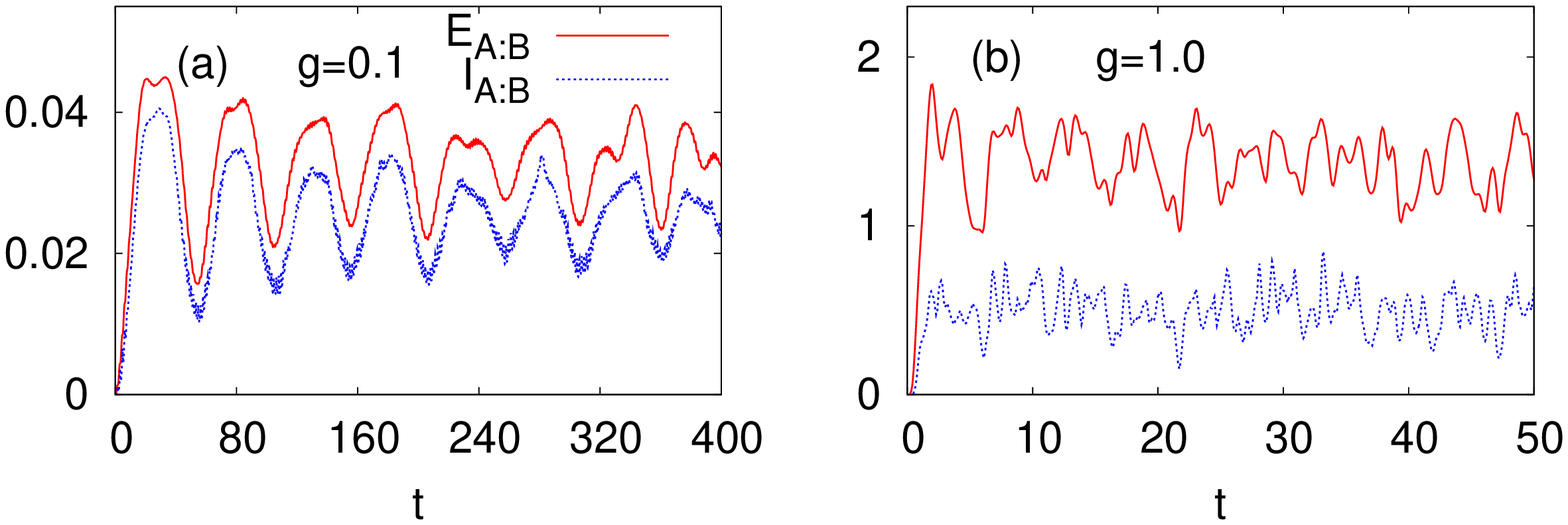}
\caption{ \label{evolution}
Entanglement $E_\sub{A:B}(L,l)$ (solid line) and mutual information 
$I_\sub{A:B}(L,l)$ (dashed line) in a time-dependent pure state of the quantum 
Ising model with $L=10$ and $l=5$. Initially all spins are aligned in the up 
direction in $z$ axis. Results for $g=0.1$ are shown in (a) and for $g=1$ in
(b).  
}
\end{figure}
%==================================

\headline{Behavior in a non-stationary pure state}
%-------------------------------------------------
The inequality $I_\sub{A:B} \leq E_\sub{A:B}$ is valid not only in the ground state
but also in time-dependent pure states of the quantum Ising chain. 
The time evolution of the density matrix is given by
\begin{equation}
\rho(t) = e^{-iHt}\vert \psi(0) \rangle \langle \psi(0) \vert e^{iHt},
\end{equation}  
where $H$ is the quantum Ising Hamiltonian~(\ref{Ham:Ising}) and
$\vert \psi(0) \rangle$ is the initial state. Since $\rho(t)$ is also
a pure state, one can obtain $E_\sub{A:B}(L,l)$ and $I_\sub{A:B}(L,l)$
using the same methods as in the ground state. 
In Fig.~\ref{evolution} we show results for $g=0.1$ and $1.0$ in a chain with 
$L=10$ and $l=5$, using an initial state that is fully magnetized in 
$z$-direction. As can be seen, the quantities oscillate irregularly but
satisfy the inequality $I_\sub{A:B} \leq E_\sub{A:B}$ at any time.

It turns out that this inequality holds generally for arbitrary local measurements on entangled pure states in bipartite system, as will be proved in the following section.

%==========================================================================
\section{Proof of the inequality $I_\sub{A:B}\leq E_\sub{A:B}$ }
%==========================================================================

To prove the inequality between mutual information and entanglement we use similar techniques as for the proof of Holevo's inequality~\cite{Holevo}. This suggests that both inequalities may be closely related or even equivalent.

\headline{Measurement}
%---------------------
In the following let us consider an arbitrary bipartite system on $\mathcal H_\sub{AB}=\mathcal H_\sub A \otimes \mathcal H_\sub B$ which is in a pure state $\rho_\sub{AB} = \vert\psi\rangle\langle\psi\vert$ so that the von-Neumann entropy $S[\rho_\sub{AB}]=-\tr\left[ \rho_\sub{AB} \ln \rho_\sub{AB} \right]$ vanishes. In such a pure state the entanglement between the subsystems $A$ and $B$ is given by
\begin{equation}
\label{standardE}
E_\sub{A:B} = S[\rho_\sub{A}]=S[\rho_\sub{B}]\,,
\end{equation}
where $\rho_\sub{A}=\tr_\sub{B}\rho_\sub{AB}$ and $\rho_\sub{B} = \tr_\sub{A}\rho_\sub{AB}$ denote the reduced density matrices. On both sides let us now perform local projective measurements 
\begin{equation}
M_\sub{A} = \sum_a a\vert \phi_a\rangle\langle \phi_a\vert = \sum_a a \Pi_a\,,\qquad
M_\sub{B} = \sum_b b\vert \phi_b\rangle\langle \phi_b\vert = \sum_b b \Pi_b
\end{equation}
with the projectors $\Pi_a=\vert \phi_a\rangle\langle \phi_a\vert$ and  $\Pi_b=\vert \phi_b\rangle\langle \phi_b\vert$, which may be thought of as spin configuration measurements as in the example given above. This measurement converts the pure state $\rho_\sub{AB}$ into a mixed state
\begin{equation}
\rho'_\sub{AB}=\sum_{a,b} \Pi_{ab} \rho_\sub{AB} \Pi_{ab}\,,
\end{equation}
where $\Pi_{ab}=\Pi_a \otimes \Pi_b$. Moreover, the measurement completely destroys the entanglement between the two subsystems. To see this, let us consider the entanglement of formation
\begin{equation}
\label{Ef}
E_F(\rho'_\sub{AB}) = \mbox{inf}\left\{ \sum_k q_k  S\Bigl[\tr_\sub{B} |k \rangle\langle k|\Bigr] \right|\left.  \rho'_\sub{AB}=\sum_k q_k |k \rangle\langle k| \right\}\,
\end{equation}
defined as the infimum of the entropy over all possible statistical ensembles represented by $\rho_\sub{AB}'$. Since $E_F(\rho'_\sub{AB})$ is non-negative and the entropy of the particular representative 
\begin{equation}
\label{measurement}
\rho_\sub{AB}'=\sum_{ab} \Pi_{ab} \; \tr \bigl[\Pi_{ab}\, \rho'\, \Pi_{ab} \bigr] 
\end{equation}
vanishes, we can conclude that $E_F(\rho'_\sub{AB})=0$ after the measurement. This means that the measurement destroys the original entanglement and converts it to some extent into classical correlations which can be quantified by the mutual information $I_\sub{A:B}$ in Eq.~(\ref{mutual}). Note that the classical mutual information $I_\sub{A:B}$ is equivalent to the quantum mutual information $I(\rho'_\sub{AB})$ of the post-measurement state $\rho'_\sub{AB}$, i.e., 
\begin{equation}
I_\sub{A:B} = I(\rho'_\sub{AB}) = S[\rho'_\sub{A}] + S[\rho'_\sub{B}] -S[\rho'_\sub{AB}]\,,
\end{equation} 
where $\rho'_\sub{A} =\tr_\sub{B}[\rho'_\sub{AB}]$ and 
$\rho'_\sub{B} =\tr_\sub{A}[\rho'_\sub{AB}]$.\\

\headline{Expressing the measurement as an isometry in an extended space}
%------------------------------------
According to the Stinespring theorem~\cite{Stinespring} the measurement process (\ref{measurement}) can be carried out by embedding $\rho_\sub{AB}$ in a higher-dimensional Hilbert space, performing a unitary transformation on it, and finally tracing out the additional degrees of freedom. To this end let us extend the Hilbert space $\mathcal H_\sub{AB} = \mathcal H_\sub A \otimes \mathcal H_\sub B$ by an auxiliary space $\mathcal H_\sub {\~A\~B}=\mathcal H_\sub{\~A} \otimes \mathcal H_\sub{\~B}$ whose task will be to store the measurement outcome encoded in the form of orthonormal basis vectors $\vert\tilde\phi_{ab}\rangle = \vert \tilde\phi_a \rangle \otimes \vert \tilde\phi_b \rangle$. Moreover, let us define the linear map 
\begin{equation}
\label{V}
V: {\mathcal H}_\sub{AB}\to {\mathcal H}_\sub{AB} \otimes \mathcal H_\sub{\~A\~B}:\;
|\psi \rangle \rightarrow
V |\psi\rangle =\sum_{ab} \langle \phi_{ab}\vert \psi\rangle\, \Bigl( \vert \phi_{ab}\rangle\otimes \vert\tilde \phi_{ab}\rangle\Bigr)\,
\end{equation}
which obeys the condition $VV^\dagger= \id$ so that it maps the initial state \textit{isometrically} onto a subset of the extended space like a unitary (entropy-preserving) transformation. With the corresponding extended density matrix
\begin{equation}
\rho_\sub {AB\~A\~B} \;:=\; V \rho_\sub{AB} V^\dagger \;=\;
\sum_{ab}\sum_{a'b'} \langle \phi_{ab}\vert \rho_\sub{AB} \vert\phi_{a'b'}\rangle\, \Bigl( \vert \phi_{ab}\rangle\langle \phi_{a'b'} \vert\otimes \vert\tilde \phi_{ab}\rangle\langle \tilde \phi_{a'b'}\vert\Bigr)\,
\end{equation}
the measurement process (\ref{measurement}) can now be written as
\begin{equation}
\rho'_\sub{AB} \;=\; \tr_\sub{\~A\~B} \bigl[\rho_\sub {AB\~A\~B}\bigr]\,.
\end{equation}
\headline{Schmidt decomposition}
%-------------------------------
%
According to Schmidt's theorem, any quantum state $\vert\psi\rangle \in {\mathcal H_\sub{AB}}$ can be decomposed into
\begin{equation}
\vert\psi\rangle \;=\; \sum_{i} \sqrt{\lambda_i}\; |i_\sub{A}\rangle\otimes |i_\sub{B}\rangle\,
\end{equation}
with certain vectors $ |i_\sub{A}\rangle\in\mathcal H_\sub{A}$, $|i_\sub{B}\rangle\in\mathcal H_\sub{B}$ and probabilities $\lambda_i\in[0,1]$, called Schmidt coefficients, which sum up to 1. This means that the initial state can be written as
\begin{equation}
\rho_\sub{AB} \;=\; |\psi\rangle\langle\psi| \;=\; \sum_{i,j} \sqrt{\lambda_i\lambda_j} \; \;
|i_\sub{A}\rangle\langle j_\sub{A}| \otimes |i_\sub{B}\rangle\langle j_\sub{B}| \,.
\end{equation}
Calculating the partial traces we obtain
\begin{equation}
\rho_\sub{A} \;=\; \tr_\sub B[\rho_\sub{AB}] \;=\; \sum_{i,j} \sqrt{\lambda_i\lambda_j} \;\;
|i_\sub{A}\rangle\langle j_\sub{A}|\;
\underbrace{\tr_\sub{B}\Bigl[|i_\sub{B}\rangle\langle j_\sub{B}|\Bigr]}_{=\delta_{ij}}
\;=\; \sum_{i} \lambda_i\, |i_\sub{A}\rangle\langle i_\sub{A}|
\end{equation}
and a similar expression for $\rho_\sub{B}$, meaning that the Shannon entropy of the Schmidt coefficients is equal to the initial entanglement:
\begin{equation}
E_\sub{A:B} = -\sum_i \lambda_i \ln \lambda_i
\end{equation}
\headline{Encoding the Schmidt decomposition in another auxiliary space}
%------------------------------------------------------------------------
%
Let us now introduce another auxiliary Hilbert space $\mathcal H_\sub C$ with canonical basis vectors $|ij\rangle$ whose task will be to store pairs of Schmidt coefficients. On the combined Hilbert space $\mathcal H_\sub{ABC} = \mathcal H_\sub{AB} \otimes \mathcal H_\sub{C}$  we define the density operator
\begin{equation}
\label{ABC}
\rho_\sub{ABC}  \;:=\;  \sum_{ij} \sqrt{\lambda_i \lambda_j} \;\;
|i_\sub{A}\rangle\langle j_\sub{A}| \;\otimes\; 
|i_\sub{B}\rangle\langle j_\sub{B}|  \;\otimes\; 
|ij\rangle\langle ij|\,.
\end{equation}
If we compare this operator with its own diagonal part
\begin{equation}
\hat\rho_\sub{ABC}  \;:=\;  \sum_{i} \lambda_i  \;\; 
|i_\sub{A}\rangle\langle i_\sub{A}| \;\otimes\; 
|i_\sub{B}\rangle\langle i_\sub{B}| \;\otimes\; 
|ii\rangle\langle ii| \,.
\end{equation}
it is easy to check that integer powers of these operators have always the same trace, i.e. $\tr[\rho_\sub{ABC}^k]=\tr[\hat\rho_\sub{ABC}^k]$ for all $k\in\mathbb N$, meaning that they have the same entropy
\begin{equation}
S[\rho_\sub{ABC}]=S[\hat\rho_\sub{ABC}]=-\sum_i \lambda_i \ln \lambda_i=E_\sub{A:B}\,.
\end{equation}
We now use the map $V$ defined in Eq.~(\ref{V}) to extend this operator even further to the space $\mathcal H_\sub {AB\~A\~BC}=\mathcal H_\sub{AB} \otimes \mathcal H_\sub{\~A\~B} \otimes \mathcal H_\sub C$ by defining
\begin{eqnarray}
\label{extended}
\rho_\sub{AB\~A\~BC} &=& 
(V\,\otimes\,\id_\sub{C})\;\rho_\sub{ABC}\;(V^\dagger\,\otimes\,\id_\sub{C})\nonumber\\[2mm]
&=& \sum_{ab}\sum_{a'b'} \sum_{ij}
\sqrt{\lambda_i \lambda_j}\;
\langle \phi_a|i_\sub{A}\rangle\langle j_\sub A|\phi_{a'}\rangle
\langle \phi_b|i_\sub B \rangle\langle j_\sub{B}|\phi_{b'}\rangle \,\, \\[-3mm]
\nonumber
&& \hspace{28mm} \times \; 
\underbrace{|\phi_{ab}\rangle\langle \phi_{a'b'}|}_{\sub{AB}} \,\otimes\,
\underbrace{|\tilde\phi_{ab}\rangle\langle\tilde \phi_{a'b'}|}_{\sub{\~A\~B}} \,\otimes\,
\underbrace{|ij\rangle\langle ij|}_{\sub{C}} \,.
\end{eqnarray}
As $V$ is an isometry, this extension does not change the entropy, hence
\begin{equation}
S[\rho_\sub{AB\~A\~BC}]=E_\sub{A:B}\,. 
\end{equation}
\headline{Tracing out the original Hilbert space $\mathcal H_\sub{AB}$ and the auxiliary space $\mathcal H_\sub{C}$ }
%------------------------------------
Tracing out the original Hilbert we have $\tr_\sub{AB}[|\phi_{ab}\rangle\langle \phi_{a'b'}|]=\delta_{aa'}\delta_{bb'}$, leading to
\begin{eqnarray}
\label{target}
\rho_\sub{\~A\~BC} &=& \tr_\sub{AB}[\rho_\sub{AB\~A\~BC}]=\sum_{ab}\sum_{ij}
\sqrt{\lambda_i \lambda_j}\;
\langle \phi_a|i_\sub{A}\rangle\langle j_\sub A|\phi_{a}\rangle
\langle \phi_b|i_\sub B \rangle\langle j_\sub{B}|\phi_{b}\rangle \,\, \\[-3mm]
&& \hspace{39mm} \times \; 
|\tilde\phi_{ab}\rangle\langle\tilde \phi_{ab}| \,\otimes\,
|ij\rangle\langle ij| \,. \nonumber
\end{eqnarray}
Since the density matrix in Eq.~(\ref{target}) has only diagonal elements,
the von-Neumann entropy of $\rho_\sub{\~A\~BC}$ and its reduced density matrices can be obtained easily as follows: 
\begin{eqnarray}
\label{E1}
S[\rho_\sub{\~A\~BC}] &=& -\sum_{i}\lambda_{i} \ln \lambda_{i} 
-\sum_{ab}\sum_{ij} \Bigl(
\sqrt{\lambda_i \lambda_j}\;
\langle \phi_a|i_\sub{A}\rangle\langle j_\sub A|\phi_{a}\rangle
\langle \phi_b|i_\sub B \rangle\langle j_\sub{B}|\phi_{b}\rangle 
\nonumber\\  && \hspace{28mm}  \times \; \ln 
\langle \phi_a|i_\sub{A}\rangle\langle j_\sub A|\phi_{a}\rangle
\langle \phi_b|i_\sub B \rangle\langle j_\sub{B}|\phi_{b}\rangle \Bigr)\,,\\
\label{E2}
S[\rho_\sub{\~AC}] &=& -\sum_{i}\lambda_{i} \ln \lambda_{i}
-\sum_{a,i} 
\lambda_i \;
\langle \phi_a|i_\sub{A}\rangle\langle i_\sub A|\phi_{a}\rangle \,
\ln \,\langle \phi_a|i_\sub{A}\rangle\langle i_\sub A|\phi_{a}\rangle\,, \\
\label{E3}
S[\rho_\sub{\~BC}] &=& -\sum_{i}\lambda_{i} \ln \lambda_{i}
-\sum_{b,i} 
\lambda_i \;
\langle \phi_b|i_\sub{B}\rangle\langle i_\sub{B}|\phi_{b}\rangle \,
\ln \,\langle \phi_b|i_\sub{B}\rangle\langle i_\sub{B}|\phi_{b}\rangle \,,
\end{eqnarray}
where 
$\rho_\sub{\~AC}=\tr_\sub{\~B}[\rho_\sub{\~A\~BC}]$ and
$\rho_\sub{\~BC}=\tr_\sub{\~A}[\rho_\sub{\~A\~BC}]$.
Combining Eqs~(\ref{E1})-(\ref{E3}), one obtains a useful relation
\begin{equation}
S[\rho_\sub{\~AC}]+S[\rho_\sub{\~BC}]-S[\rho_\sub{\~A\~BC}] = E_\sub{A:B} \,.
\end{equation}
Tracing out the auxiliary space $\mathcal H_\sub C$ we obtain the reduced density matrices $\rho_\sub{\~A\~B},\rho_\sub{\~A}$, and $\rho_\sub{\~B}$ in which the outcome of the measurement is stored. Therefore, the classical mutual information after the measurement can be expressed in terms of these density matrices by
\begin{equation}
S[\rho_\sub{\~A}]+S[\rho_\sub{\~B}]-S[\rho_\sub{\~A\~B}] =I_\sub{A:B}\,. 
\end{equation}

\headline{Apply strong subadditivity}
%------------------------------------
%
To prove the inequality $I_\sub{A:B}\leq E_\sub{A:B}$, we use the strong subadditivity of the von-Neumann entropy~\cite{SubAdd} which is known to hold in the two following equivalent forms~\cite{Nielsen}:
\begin{eqnarray}
\label{SA1}
S[\rho_\sub{\~A\~BC}]+S[\rho_\sub{\~B}]&\leq & S[\rho_\sub{\~A\~B}]+S[\rho_\sub{\~BC}] \\
\label{SA2}
S[\rho_\sub{\~A}]+S[\rho_\sub{\~B}]&\leq & S[\rho_\sub{\~AC}]+S[\rho_\sub{\~BC}]\,.
\end{eqnarray}
Using the first relation the mutual information can be written as
\begin{equation}
\label{I1}
I_\sub{A:B} \;=\; S[\rho_\sub{\~A}]+S[\rho_\sub{\~B}]-S[\rho_\sub{\~A\~B}]  \;\leq \;         S[\rho_\sub{\~A}]+S[\rho_\sub{\~BC}]-S[\rho_\sub{\~A\~BC}]
\end{equation}
Replacing $\tilde A\leftrightarrow \tilde B$ one obtains in the same way
\begin{equation}
\label{I2}
I_\sub{A:B} \;\leq \;  S[\rho_\sub{\~B}]+S[\rho_\sub{\~AC}]-S[\rho_\sub{\~A\~BC}]\,.
\end{equation}
Adding (\ref{I1}) and (\ref{I2}) gives an inequality where we again apply the second version of strong subadditivity~(\ref{SA2})
\begin{eqnarray}
2I_\sub{A:B} &\leq&  S[\rho_\sub{\~A}]+S[\rho_\sub{\~B}]+S[\rho_\sub{\~AC}]+S[\rho_\sub{\~BC}] -2S[\rho_\sub{\~A\~BC}] \nonumber \\ &\leq& 2S[\rho_\sub{\~AC}]+2S[\rho_\sub{\~BC}] -2S[\rho_\sub{\~A\~BC}]\;=\; 2E_\sub{A:B}\,,
\end{eqnarray}
proving the initial assertion.

%==========================================================================
\section{Conclusion}
%==========================================================================

We have investigated the relation between the initial entanglement and 
the classical mutual information after local projective measurements in a bipartite
quantum system which is initially in a pure state. For the quantum Ising chain with
periodic boundary conditions, which is divided into two segments, it is found 
that both quantities obey the same scaling form at the critical point, differing only
in the prefactor of the scaling function. For the entanglement we find the prefactor
$c/3=1/6$ in agreement with predictions of conformal field theory, whereas  
the prefactor of the mutual information is found to be less than $1/6$.

Furthermore, we have observed that classical mutual information cannot exceed 
the initial entanglement in arbitrary pure states. The inequality $I_\sub{A:B}
\leq E_\sub{A:B}$ has been proved generally by successively expansions and reducing
the Hilbert space and by applying strong subadditivity of the von-Neumann entropy. 
For entangled pure states, we conclude that a local projective measurement destroy the 
original quantum correlation and converts it into classical mutual information 
bounded from above by the initial entanglement.
  
In a general case, it is more convenient using the quantum mutual information
$I(\rho_\sub{AB}) = S[\rho_\sub{A}] + S[\rho_\sub{B}] -S[\rho_\sub{AB}]$
to measure the correlation in a quantum system. According to the monotonicity
of the quantum relative entropy, the local projective measurement does not
increase the quantum mutual information~\cite{monotonicity}, that is,
\begin{equation}
\label{qmutual}
I(\rho'_\sub{AB})=S[\rho'_\sub{A}] +S[\rho'_\sub{B}] -S[\rho'_\sub{AB}]
\leq I(\rho_\sub{AB}),
\end{equation}
where $\rho'_\sub{AB}$ is the post measurement density operator.
If the initial system is in a pure state, Eq.~(\ref{qmutual}) yields
$I_\sub{A:B} \leq 2 E_\sub{A:B}$ which is consistent with our
inequality. As a future work, generalized inequalities induced by the 
measurement on mixed state will be studied in a context of the quantum
mutual information.  

\ack
This work was supported by the Basic Science Research Program of MEST, NRF grant No.2010-0009697.


\section*{References}

\begin{thebibliography}{99}

\bibitem{CI}
Henkel M 1999 
\textit{Conformal Invariance and Critical Phenomena} (Berlin: Springer)

\bibitem{Vidal}
Vidal G, Latorre J I, Rico E and Kitaev A 2003
\textit{Phys. Rev. Lett.} {\bf 90} 227902

\bibitem{Holzhey} 
Holzhey C, Larsen F and Wilczek F 1994
\textit{Nucl. Phys.} B {\bf 424} 443

\bibitem{Calabrese1}
Calabrese P and Cardy J 2004
\textit{J. Stat. Mech.} P06002
\bibitem{Calabrese2}
Calabrese P and Cardy J 2006
\textit{Int. J. Quant. Inf.} {\bf 4} 429 
\bibitem{Calabrese3}
Calabrese P and Cardy J 2009
\textit{J. Phys. A} {\bf 42} 504005


\bibitem{lanczos}
Dagotto E1994
\textit{Rev. Mod. Phys.} {\bf 66} 763

\bibitem{Stinespring}
Stinespring F 1955 
\textit{Proc. Amer. Math. Soc} 211

\bibitem{Holevo}
Holevo A S 1973 
\textit{Teor. Veroyatnost. i Primenen.} \textbf{18} 371 

\bibitem{SubAdd}
Lieb E H and Ruskai M B 1973 
\textit{J. of M. Phys.} \textbf{14} 1938

\bibitem{Nielsen}
Nielsen M A and Chuang I L 2000 
\textit{Quantum Information and Quantum Computation} (Cambridge: Cambridge University Press) p~521

\bibitem{monotonicity}
Sagawa T 2012 
\textit{Lectures On Quantum Computing, Thermodynamics And Statistical Physics} 
(\textit{Kinki University Series on Quantum Computing}) ed M Nakahara and S Tanaka (World Scientific), \textit{Preprint} cond-mat.stat-mech/12020983 

 
\end{thebibliography}
\end{document}